\documentclass{JHEP3}
\relax

\def\bes{\begin{eqnarray}}
 \def\ees{\end{eqnarray}}
\def\be{\begin{equation}}
\def\ee{\end{equation}}
\def\bs{\begin{subequations}}
\def\es{\end{subequations}}
\newcommand{\een}{\end{subequations}}
\newcommand{\ben}{\begin{subequations}}
\newcommand{\beq}{\begin{eqalignno}}
\newcommand{\eeq}{\end{eqalignno}}

\def\rt{{\tilde{r}}}
\def\ti{{\tilde{t}}}
\def\phit{{\tilde{\phi}}}
\def\ct{{\tilde{c}}}
\def\Lt{{\tilde{L}}}
\def\Dt{{\tilde{\Delta}}}
\def\chit{{\tilde{\chi}}}

\title{Entropy from AdS$_3$/CFT$_2$}

\author{N. Tetradis
\\
Department of Physics, University of Athens, Zographou 157 84, Greece 

\\
{\tt ntetrad@phys.uoa.gr}}

\preprint{}

\abstract{
 We parametrize the (2+1)-dimensional AdS space and the BTZ black hole with 
Fefferman-Graham coordinates starting from the AdS boundary. We consider various boundary metrics:
Rindler, static de Sitter and FRW. In each case, 
we compute the holographic stress-energy
tensor of the dual CFT and confirm that it has the correct form, including the effects of the conformal anomaly.
We find that the Fefferman-Graham parametrization also spans a second copy of the AdS space, including a second boundary.
For the boundary metrics we consider, the Fefferman-Graham coordinates do not cover the whole AdS space. 
We propose that the length of the line delimiting the excluded region at a given time can be identified with the entropy of the dual CFT 
on a background determined by the boundary metric.
For Rindler and de Sitter backgrounds our proposal reproduces the expected entropy. For a
FRW background it produces a generalization of the Cardy formula that takes into account the
vacuum energy related to the expansion. 
}


\begin{document}

\section{Introduction}
\label{intrr}
At low-energies the AdS/CFT correspondence provides a connection between supergravity on 
an asymptotically anti-de Sitter (AdS) bulk space and a conformal field theory (CFT) that can be considered
as living on its boundary \cite{adscft1,adscft2}.
Even though the original proposal \cite{adscft1} concerns the 
$\mathcal{N}=4$ super Yang-Mills theory with $SU(N)$ gauge symmetry in the large-$N$
limit in four dimensions, the correspondence is assumed to be applicable to CFTs on spaces of various 
dimensionalities. Our interest lies in the thermodynamic properties of such theories, especially when the
background geometry on which they are assumed to exist is nontrivial. The methodology for
addressing this issue can be based on well established principles. 
The properties of the CFT at nonzero temperature can be deduced from a bulk geometry that 
contains a black hole \cite{witten}. In the simplest example, a (4+1)-dimensional AdS black hole encodes
information for the  $\mathcal{N}=4$ super Yang-Mills theory at nonzero temperature.
Thermodynamic quantities, such as the energy density, can be derived 
through the calculation of the stress-energy tensor of the dual CFT, after the appropriate
renormalization procedure has been carried out \cite{se,Skenderis}. The boundary metric
is not dynamical in this procedure, but acts as a source for the stress-energy tensor. Even though
it is usually taken to be flat, it can be quite general (assuming that no pathologies are encountered
in the reconstruction of the bulk space-time with such a boundary condition).

We are mainly interested in the entropy of boundary CFTs on nontrivial gravitational backgrounds. In order to
address the problem through holographic methods, it is important to identify holographic screens that may extend
deep into the bulk. As a result, a solution for the full bulk metric is necessary, not just its expansion near the
boundary. The problem is simplified if the boundary metric belongs
to the same conformal class as a simpler metric for which the bulk extension is known.
In such a case, an appropriate bulk coordinate transformation can generate the required form of the boundary metric. 
An example of this procedure is given in \cite{ast,tetradis}.  (See also ref. \cite{kajantie}, as well
as ref. \cite{meyer} for a generalization with a bulk dilaton.) The (4+1)-dimensional AdS black hole 
solution (with a spherical, flat, or hyperbolic horizon) is expressed in coordinates such that the boundary metric is 
of the Friedmann-Robertson-Walker (FRW)
type. The stress-energy tensor of the dual thermalized CFT on an expanding background can be calculated through 
holographic renormalization \cite{Skenderis}. Its expected form is reproduced correctly, including the effects of the
conformal anomaly. Moreover, the entropy of the CFT can be studied through the bulk geometry.

In this work, we repeat the same procedure for (2+1)-dimensional
AdS space, which becomes the (2+1)-dimensional BTZ black hole through appropriate 
identifications \cite{btz1,btz2,carlip}. We cast the bulk metric in the Fefferman-Graham form \cite{fg} near the
boundary. 
We also expand the scope of the study by considering boundary metrics of the Rindler and de Sitter type. 
The simpler form of the (2+1)-dimensional AdS space permits the extraction of exact expressions for the 
coordinate transformations that lead to various forms of the boundary metric. In some cases we find
more than one bulk metrics with the same boundary, which correspond to various states of the dual CFT.
Using the Fefferman-Graham parametrization, we derive the holographic stress-energy tensor and the conformal anomaly 
in each case.

Another important consequence of the use of Fefferman-Graham coordinates is that the global properties of space 
can be altered, even though locally the metric is always isometric to AdS. We find that, for the boundary metrics of 
interest, the coordinates also span a second copy of AdS, including a second boundary. The two copies of AdS are joined
through a surface that may resemble a throat or bridge,  similar to the Einstein-Rosen bridge of the
four-dimensional Schwarzschild black hole \cite{mtw}, or have a more complicated structure. Moreover, portions of
AdS may be missing from either copy.

With respect to the entropy, we search for appropriate holographic screens, whose
area can be identified with the entropy of the dual CFT.
For a Rindler or FRW boundary metric, the narrowest part of the throat or bridge at a given time delimits the region not covered 
by the Fefferman-Graham parametrization.
We show that its length 
can be identified with the entropy of the dual CFT on the respective background. For a static de Sitter boundary, the 
global structure of space is rather complicated. However, again the entropy can be identified with the length of 
the line delimiting the region of AdS not covered by the Fefferman-Graham coordinates.
The calculation of the entropy is delicate for the Rindler and de Sitter boundaries, as the boundary metric is not dynamical and the
two-dimensional Newton's constant vanishes. Through an appropriate regularization of the near-boundary bulk space-time, we
show that the entropy of two-dimensional Rindler and de Sitter spaces is reproduced correctly. For a boundary of the FRW type, 
the throat becomes time-dependent. We discuss the corresponding entropy and its interpretation in terms of a generalization of
the Cardy formula \cite{cardy}.

The entropy of two-dimensional de Sitter space has also been computed through holography in \cite{dsen,iwashita}
by means of the Randall-Sundrum construction \cite{rs}. In contrast to these works, we do not assume the presence
of a UV cutoff for the CFT.

 The entaglement entropy of the dual CFT within a specified region ${\cal A}$ on the boundary 
has been studied through a holographic approach in \cite{ryu,takayanagi,hubeny}. The entropy is identified with the area of 
an appropriately defined minimal surface ${\cal S}$ that starts from the border of ${\cal A}$ on the boundary and 
extends into the bulk. When ${\cal A}$ is taken to cover the whole boundary,  ${\cal S}$ tends to wrap around
obstructions in the bulk space-time, such as black-hole horizons. In this limit and for
flat boundaries, the approach of \cite{ryu,takayanagi,hubeny} reproduces the thermodynamic entropy of the 
dual CFT. This proposal is very similar in spirit to our approach. The various throats or bridges that we shall discuss
provide obstructions around which a minimal surface would wrap. 

In section \ref{3db} we review standard coordinate systems for the (2+1)-dimensional AdS space and the BTZ black hole. 
In section \ref{mink}
we discuss the AdS/CFT correspondence for a bulk BTZ black hole with a static boundary. 
In section \ref{rindd} we present two metrics describing AdS space with a Rindler boundary. The first corresponds to
the ground state of the dual CFT, while the second to an excited state. In \ref{sdsb} we do the same for a 
static de Sitter boundary. In section \ref{expbound} we consider the BTZ black hole with 
a time-dependent boundary of the FRW type. In section \ref{lthr} we present our proposal for the entropy and
test it on the various backgrounds. In section \ref{conclu} we give our conclusions.

\section{(2+1)-dimensional AdS space and the BTZ black hole}
\label{3db}

We consider solutions of the Einstein field equations in 2+1 dimensions with a 
negative cosmological constant $\Lambda_3 = -1/l^2$.  All such solutions are locally isometric to AdS space. 
We set $l =1$ for simplicity throughout the paper.

\subsection{BTZ black hole}
The metric for the non-rotating BTZ black hole \cite{btz1} can be written in Schwarzschild coordinates as
\begin{equation}
\label{eqmetric} 
ds^2 = -f(r) dt^2 + \frac{dr^2}{f(r)} + r^2 d\phi^2,\ \ \ \ \ \ f(r) = r^2- \mu. 
\end{equation}
 In order for the above solution to have the properties of a black hole of mass $\sim \mu$, the coordinate
$\phi$ must be periodic, with period equal to $2\pi$. 
The Hawking temperature, energy (or mass) and entropy of the black hole are, respectively,
\begin{equation}
\label{eqTM}
 T = \frac{1}{2\pi}\sqrt{\mu}, 
\ \ \ \ \ \ \ 
E = \frac{V}{16\pi G_3}\mu,
\ \ \ \ \ \ \ 
S = \frac{V}{4 G_3}\sqrt{\mu},
\end{equation}
with $V$ the volume of the one-dimensional space spanned by $\phi$ (i.e. $V=2\pi$), 
and $G_3$ Newton's constant.

\subsection{AdS in Poincare coordinates}
If $\mu$ is assumed to vanish and the periodicity of $\phi$ by $2\pi$ is lifted, 
eq. (\ref{eqmetric}) describes part of the covering space of AdS.  
This becomes obvious if we observe that, 
for $\mu=0$ and $\phi$ taking values over the
whole real axis,  the definition of a new coordinate $u=1/r$ turns eq. (\ref{eqmetric}) into the standard 
AdS metric in Poincare coordinates. 

\subsection{AdS in global coordinates}
\label{globalcoord}
In the following we shall find very useful the parametrization of AdS space in
terms of global coordinates. The form of the metric, which covers the whole AdS space, is 
\begin{equation}
ds^2=-\cosh^2(\rt) d\ti^2 +d\rt^2+\sinh^2(\rt)d\phit^2.
\label{global} \end{equation}
For $0<\ti<2 \pi$, eq. (\ref{global}) describes AdS space. If $\ti$ is allowed to take values over the whole
real axis, we obtain the covering space of AdS. The boundary of AdS is approached for $\rt \to \infty$.
If we define a coordinate $\chit$ through $\tan(\chit)=\sinh(\rt)$, the metric
(\ref{global}) becomes
\begin{equation}
ds^2=\frac{1}{\cos^2(\chit)} \left[ -d\ti^2 +d\chit^2+\sin^2(\chit)\, d\phit^2 \right].
\label{global2} \end{equation}
The new coordinate takes values $0\leq \chit < \pi/2$. 
The boundary is now approached for $\chit \to \pi/2$.
The relation between
Poincare and global coordinates is \cite{review}
\begin{eqnarray}
\ti(t,r,\phi)&=&\arctan \left[ \frac{2r^2 t}{1+r^2(1+\phi^2-t^2)}\right]
\label{globalt} \\
\chit(t,r,\phi)&=&\arctan \sqrt{r^2\phi^2+\frac{\left[ 1-r^2(1-\phi^2+t^2) \right]^2}{4r^2}}
\label{globalr} \\
\phit(t,r,\phi)&=&\arctan \left[ \frac{1-r^2(1-\phi^2+t^2)}{2r^2 \phi} \right].
\label{globalf} \end{eqnarray}
As the spatial part of the AdS boundary is compact, the global coordinate $\phit$ is
periodic with period $2\pi$. It is obvious then from eq. (\ref{globalf}) that the limits $\phi\to\pm\infty$ of the 
Poincare coordinate $\phi$ (for $\mu=0$ in eq. (\ref{eqmetric})) must be identified.


\subsection{Other local parametrizations}
For $\mu\not=0$ and $\phi$ periodic, with period $2\pi$, the metric (\ref{eqmetric}) describes the BTZ black hole.
The relation between the Schwarzschild and global 
coordinates is given in ref. \cite{carlip}, but we shall not use it in this work. 
For any value of $\mu$ the metric
(\ref{eqmetric}) is locally isometric to AdS, independently of the range of values of $\phi$.
The physical interpretation of this metric with $\mu\not= 0$ and $\phi$ taking values over the whole real axis
is not clear in the context of the AdS/CFT correspondence.
Despite that,  in the following
we consider solutions with nonzero $\mu$, even when 
we do not identify them with the BTZ black hole. 
For $\mu\not=0$ we refer to $\phi$ as a periodic
coordinate, with period $2\pi$, when the metric describes the BTZ black hole, and as non-periodic when we allow it
to take values over the whole real axis. We always assume that $\mu \geq 0$.

\section{Minkowski boundary}
\label{mink}
In order to set the framework for the following sections, we first review the dual interpretation of the BTZ 
black hole through the AdS/CFT correspondence, employing Fefferman-Graham coordinates \cite{fg}. 
We start by expressing the metric (\ref{eqmetric}) in 
terms of such coordinates. This can be achieved by defining a new coordinate $z$ through
${dz}/{z} = - {dr}/{\sqrt{f(r)}}. $
We obtain
\begin{equation}
z=\frac{2}{\mu}\left(r-\sqrt{r^2-\mu} \right),
\label{zofrm}
\end{equation}
where we have chosen the multiplicative constant so that $z\simeq 1/r$ for $r\to \infty$.
Inverting this relation we find 
\begin{equation}
r=\frac{1}{z}+\frac{\mu}{4}z.
\label{rofz}
\end{equation}
The coordinate $z$ takes values in the interval $[0,z_e=2/\sqrt{\mu}]$, covering the region outside the event
horizon in which $r$ takes values in the interval $[r_e=\sqrt{\mu},\infty]$.
The metric of eq. (\ref{eqmetric}) takes the form
\begin{equation}
\label{eqmetric1} 
ds^2 = \frac{1}{z^2} 
\left[ dz^2 - \left( 1-\frac{\mu}{4}z^2\right)^2 dt^2 +\left( 1+\frac{\mu}{4}z^2\right)^2 d\phi^2 \right]. 
\end{equation}
Eq. (\ref{rofz}) is also satisfied for 
\begin{equation}
z=\frac{2}{\mu}\left(r+\sqrt{r^2-\mu} \right).
\label{zofrp}
\end{equation}
This expression results from the condition
${dz}/{z} = + {dr}/{\sqrt{f(r)}}. $
The metric of eq. (\ref{eqmetric}) takes again the form
of eq. (\ref{eqmetric1}), but now the coordinate $z$ takes values 
in the interval $[z_e=2/\sqrt{\mu},\infty]$, covering again the region outside the event
horizon in which $r$ takes values in the interval $[r_e=\sqrt{\mu},\infty]$.

We may consider the metric of eq. (\ref{eqmetric1}) allowing $z$ to vary in the interval $[0,\infty]$.
In this case the region outside the horizon is covered twice. 
The coordinates $(t,z)$ are similar to the isotropic coordinates that are 
often employed for the study of the Schwarzschild geometry.
The isotropic coordinates do not span the full space. 
They cover the two regions of the Kruskal-Szekeres plane that are located
outside the horizons.  For fixed coordinate time, the metric describes a throat
connecting two asymptotically flat regions \cite{mtw}. 
A similar situation occurs for the coordinates $(t,z)$ in the case of the BTZ black hole in 2+1 dimensions, or
the AdS-Schwazschild geometry in higher dimensions \cite{tetradis}. In the following we shall
also find generalizations of this construction for various boundary metrics.

An asymptotically AdS geometry can be related 
to a dual CFT on the AdS boundary through the AdS/CFT correspondence.
 The most general (2+1)-dimensional metric that satisfies Einstein's equations with 
a negative cosmological constant is of the form \cite{finite}
\be\label{eq2} ds^2 = \frac{1}{z^2} \left[ dz^2 + g_{\mu\nu} dx^\mu dx^\nu \right], \ee
where
\be g_{\mu\nu} = g_{\mu\nu}^{(0)} + z^2 g_{\mu\nu}^{(2)} 
+ z^4 g_{\mu\nu}^{(4)}. \label{fefg} \ee
(However, the global properties of the geometry can be quite nontrivial \cite{global}.)
The stress-energy tensor of the dual CFT is \cite{Skenderis}
\begin{equation}
\label{eq3a} 
\langle T_{\mu\nu}^{(CFT)} \rangle =
\frac{1}{8\pi G_3} 
\left[ g^{(2)}-{\rm tr}\left( g^{(2)}\right)g^{(0)} \right].
\end{equation}

We apply this general expression to the metric (\ref{eqmetric1}) of the BTZ black hole, which has
a flat boundary
\begin{equation}
\label{eqmetric0} 
ds_0^2 = g_{\mu\nu}^{(0)} dx^\mu dx^\nu = -dt^2 +d\phi^2.
\end{equation}
We obtain the energy density and pressure, respectively,
\begin{eqnarray}
\label{eq3e} 
\rho&=&\frac{E}{V}=-\langle T_{~t}^{t} \rangle=\frac{\mu}{16\pi G_3},
\\
p& =& \langle T_{~\phi}^{\phi} \rangle = 
\frac{\mu}{16\pi G_3},
\label{eq3p}  
\end{eqnarray}
on a boundary with metric (\ref{eqmetric0}).


The Hawking temperature $T$ of the black hole \cite{page} can be
determined from the metric of eq. (\ref{eqmetric1}). In Euclidean space, 
the metric possesses a conical singularity at the location of the horizon
$z_e= 2/\sqrt{\mu}$. This can be eliminated if the Euclidean time is 
periodic, with period $1/T$. Expanding $z$ around $z_e$, we find that the conical
singularity disappears for $T$ taking the value 
given by the first of eqs. (\ref{eqTM}).
In order to determine the entropy, we consider a variation of the 
parameter $\mu$ of the metric (\ref{eqmetric1}). This variation does not 
affect the volume $V$ of the boundary. The variations of the energy density 
$E$ and entropy $S$ obey $dE=TdS$. A simple integration returns the expression for the entropy given in the third 
of eqs. (\ref{eqTM}).
As expected, the entropy is proportional to the surface of the event horizon. For the
parametrization of eq. (\ref{eqmetric1}), the event horizon coincides with the narrowest part of the throat.

In the dual picture, the solution (\ref{eqmetric1}) can be interpreted as a CFT at a nonzero temperature, equal to the
Hawking temperature of the black hole. The energy density is given by eq. (\ref{eq3e}), so that the total energy
is proportional to the black hole mass. The entropy is also equal to that of the black hole. The mass term $\mu$ sets the scale
of the temperature relative to the AdS length $\ell=1$. We assume that $\mu\gg 1$.

In 2+1 dimensions the Fefferman-Graham expansion terminates with the term $\sim z^4$, as can be seen from eq. (\ref{fefg}). 
(This simplification in general does not occur in higher dimensions.) 
As a result, we can obtain information on the dual CFT on a different background 
corresponding to the limit $z\to \infty$. It can 
be checked easily that the metric  (\ref{eq2}) retains the Fefferman-Graham form when expressed in terms of $z'=1/z$. 
The AdS boundary is now located at $z'=0$. The boundary metric is $ g_{\mu\nu}^{(4)}$, which 
in general is different than the metric  $g_{\mu\nu}^{(0)}$, relevant for the boundary at $z=0$. 
For the metric (\ref{eqmetric1}) the construction we described in this section joins a portion of the AdS space, from the boundary
at $z=0$ with metric $g_{\mu\nu}^{(0)}=\eta_{\mu\nu}$ up to the black hole horizon, with another portion of AdS, from the 
black hole horizon up to the boundary at $z'=0$ with metric $g_{\mu\nu}^{(4)}=(\mu^2/16)\eta_{\mu\nu}$.
Each copy of (part of) the AdS space can be considered as a holographic dual of the CFT on the respective boundary.
In this example both boundaries are flat, but we shall see examples with different boundary metrics in the following.

\section{Rindler boundary}
\label{rindd}
\subsection{Ground state}
\label{gsrindd}
We would like to repeat the construction of the previous section for a Rindler boundary metric.
For the Rindler wedge ($x>0$), the metric (\ref{eqmetric}) can be put in the form
\begin{equation}
ds^2
= \frac{1}{z^2} \left[ dz^2 
- a^2 x^2 \left(1-\frac{1}{4}\left[\frac{\mu-a^2}{a^2}\right]\frac{z^2}{x^2} \right)^2 dt^2 
+  \left(1+\frac{1}{4}\left[\frac{\mu-a^2}{a^2}\right]\frac{z^2}{x^2} \right)^2 dx^2 \right].
\label{rindb}
\end{equation}
The boundary metric
\begin{equation}
\label{rind} 
ds_0^2 = g_{\mu\nu}^{(0)} dx^\mu dx^\nu = -a^2 x^2 dt^2 + dx^2.
\end{equation}
is of the two-dimensional Rindler form.

The coordinate transformation that results in (\ref{rindb}) does not affect the time coordinate. It is given by
\begin{eqnarray}
r(z,x)&=&\sqrt{\mu+\frac{a^2x^2}{z^2}\left[ 1-\frac{\mu-a^2}{a^2}\frac{z^2}{4x^2}\right]^2 }
\label{brind2} \\
\phi(z,\rho)&=&\frac{1}{a}\log\left[ a x \right]
+\frac{1}{2\sqrt{\mu}}
\log\left[\frac{4a^2\left(\sqrt{\mu}-a\right)^2+(\mu-a^2)^2z^2/x^2}{4a^2\left(\sqrt{\mu}+a\right)^2+(\mu-a^2)^2z^2/x^2} \right].
\label{brind3a} 
\end{eqnarray}
Notice that the transformation (\ref{brind3a}) maps the whole region near negative infinity 
for $\phi$ to the neighborhood of zero for $x$.

For fixed $x$,  $r(z,x)$ has a minimal value as a function of $z$, obtained at the point where
$\partial r/\partial z=0$. Also, for fixed $z$, there is a point at which the Fefferman-Graham parameters start covering the
AdS space for a second time. This point corresponds to $\partial r/\partial x=0$. For the case of a Rindler boundary, both
conditions define the same line on a slice of constant $t$. 
For $\mu>a^2$ the line corresponds to 
$z_m^2(x)=4a^2x^2/(\mu-a^2)$, on which $r_m(x)=\sqrt{\mu}$, while for $\mu<a^2$ it corresponds to
$z_m^2(x)=4a^2x^2/(-\mu+a^2)$, on which $r_m(x)=a$. 
This line defines the narrowest part of a bridge connecting the
two asymptotic regions at $z\to 0$ and $z\to \infty$. We shall return to this point in the
discussion of the entropy. 

It is also apparent from the above expressions that, if we would like to have $x$ take values in the region $[0,\infty]$, the 
Schwarzschild variable $\phi$ of eq. (\ref{eqmetric}) cannot be periodic, with a finite period. 
Instead it must take values over the whole real axis. As a result, the
metric of eq. (\ref{rindb}) does not represent the BTZ black hole, but (part of) the covering space of AdS. A natural physical
interpretation can be given only to the case $\mu=0$, for which eq. (\ref{eqmetric}) is equivalent to the Poincare parametrization
of AdS (through the definition of a new variable $u=1/r$). As we discussed in section \ref{3db}, the limits $\phi\to\pm \infty$ must
be identified in this case, which means that the limits $x\to 0$ and $x\to \infty$ must be identified as well. This fact has 
important implications for the calculation of the entropy.

As in the following we shall discuss in detail the case $\mu=0$, 
we give explicit expressions for the quantities of interest.
For $\mu\to 0^+$ we obtain 
\begin{eqnarray}
r(z,x)&=&a\left(\frac{x}{z}+\frac{z}{4x} \right)
\label{bbb22} \\
\phi(z,x)&=&\frac{1}{a}\log\left[a x \right]-\frac{8}{a(4+z^2/x^2)}.
\label{bbb33}
\end{eqnarray}
The minimal value $r_m(x)$ and the corresponding values of $z$ and $\phi$ are given by  
\begin{eqnarray}
z_m(x)&=&2x
\label{zmx} \\
r_m(x)&=&a
\label{rmx} \\
\phi_m(x)&\equiv&\phi_m(z_m(x),x)=(\log[ax]-1)/a.
\label{phimx} 
\end{eqnarray}

Applying the expression (\ref{eq3a}) to the metric of eq. (\ref{rindb}) we obtain the energy density and pressure, respectively,
\begin{eqnarray}
\label{eq3r} 
\rho&=&-\langle T_{~t}^{t} \rangle=\frac{1}{16\pi G_3}\frac{\mu-a^2}{a^2x^2},
\\
p& =& \langle T_{~x}^{x} \rangle = \frac{1}{16\pi G_3}\frac{\mu-a^2}{a^2x^2},
\label{eq3s}  
\end{eqnarray}
on a boundary with metric (\ref{rind}). The stress-energy tensor depends on the 
parameter $\mu$, which introduces an arbitrary energy scale in the  problem. However, we have seen 
already that the value $\mu=0$ must be selected in order for eq. (\ref{eqmetric}) to represent AdS space in Poincare coordinates.
This means that it is natural to
identify the ground state of the CFT with the solution with $\mu=0$.  
The stress-energy tensor displays the expected singularity at the location 
of the horizon at $x=0$ \cite{bd}.
The conformal anomaly vanishes.

As we discussed in the previous section, the AdS boundary is also approached in the limit $z\to \infty$. 
As a result, the metric (\ref{rindb}) can be viewed as the holographic dual of a CFT on a background determined by
the term $g^{(4)}_{\mu\nu}$ in the expansion (\ref{fefg}). It can be verified, through the definition of the new coordinates
$z'=1/z$ and $x'=1/x$, that the boundary metric is again of the Rindler type.
For the metric (\ref{rindb}) with $\mu=0$ the construction we described in this section joins a portion of the AdS space, from the boundary
at $z=0$ with a Rindler metric up to the surface with $r=a$ (see eq. (\ref{rmx})), with another portion of AdS, from the 
surface $r=a$ up to the boundary at $z'=0$ with a second Rindler metric.
Each copy of (part of) the AdS space can be considered as a holographic dual of the CFT on the respective boundary.

\subsection{Excited state}
\label{esrindd}
It is interesting to note that the form of the metric (\ref{rindb}) is not the only one with a Rindler boundary.
The metric (\ref{eqmetric}) with $\mu=0$ can be put in the form
\begin{equation}
ds^2
= \frac{1}{z^2} \left[ dz^2 
- a^2 x^2 d\eta^2 +  dx^2 \right],
\label{rindbb}
\end{equation}
with a boundary again given by 
(\ref{rindb}).
The coordinate transformation that achieves this is given by
\begin{eqnarray}
t(\eta,x)&=&x \sinh(a\eta)
\\
r(z)&=&\frac{1}{z}
\label{brind22} \\
\phi(z,x)&=&x \cosh(a\eta) .
\label{brind33}
\end{eqnarray}

We interpret this solution as being dual to an excited state of the CFT on a Rindler boundary. This can be deduced from the 
fact that 
the corresponding stress-energy tensor vanishes. On the contrary, the energy density of eq. (\ref{eq3r})
becomes negative for $\mu=0$, and vanishes only if we assume the presence of a positive energy scale $\mu=a^2$.

\section{Static de Sitter boundary}
\label{sdsb}
\subsection{Ground state}
\label{gssdsb}
We now turn to the case of a de Sitter boundary in static coordinates.
The metric (\ref{eqmetric}) can be put in the form
\begin{eqnarray}
ds^2
= \frac{1}{z^2} \Bigg[ dz^2 
&-& (1-H^2\rho^2)\left(1-\frac{1}{4}\left[\frac{\mu-H^2}{1-H^2\rho^2}+H^2\right]z^2 \right)^2 dt^2 
\Biggr.
\nonumber \\
\Biggl.
&+& \left(1+\frac{1}{4}\left[\frac{\mu-H^2}{1-H^2\rho^2}-H^2\right]z^2 \right)^2  \frac{d\rho^2}{1-H^2\rho^2} \Biggr],
\label{eqmetric4} \end{eqnarray}
with a de Sitter boundary metric   
\begin{equation}
\label{eqmetric30} 
ds_0^2 = g_{\mu\nu}^{(0)} dx^\mu dx^\nu = -(1-H^2 \rho^2) dt^2 + \frac{d\rho^2}{1-H^2\rho^2}
\end{equation}
in static form.
We concentrate on the causally connected region $-1/H < \rho < 1/H$. 

The coordinate transformation does not affect the time coordinate. It is given by 
\begin{eqnarray}
r(z,\rho)&=&\sqrt{\mu+\frac{\left[ 1-H^2\rho^2-(\mu-H^4\rho^2)z^2/4\right]^2}{z^2(1-H^2\rho^2)} }
\label{bb2} \\
\phi(z,\rho)&=&\frac{1}{2H}\log\left[\frac{1+H\rho}{1-H\rho} \right]
+\frac{1}{2\sqrt{\mu}}\log\left[\frac{1-H^2\rho^2+\left(\sqrt{\mu}-H^2\rho \right)^2z^2/4}{1-H^2\rho^2+\left(\sqrt{\mu}+H^2\rho \right)^2z^2/4} \right].
\label{bb3a}
\end{eqnarray}
 It is noteworthy that the transformation 
maps the region near negative infinity for $\phi$ to the vicinity of $-1/H$ for $\rho$, and the region near positive infinity for
$\phi$ to the vicinity of $1/H$ for $\rho$.

Similarly to the Rindler case, at a given time and 
for fixed $\rho$ there is a minimal value for $r(z,\rho)$ as a function of $z$. This value is
obtained for $z_m^2(\rho)=4(1-H^2\rho^2)/|H^4\rho^2-\mu|$. On this line we have
$r_m(\rho)=H^2|\rho|$ for $H^2 |\rho|>\sqrt{\mu}$ , and $r_m(\rho)=\sqrt{\mu}$ for  $H^2 |\rho|<\sqrt{\mu}$.
Also, for fixed $z$ there is a point at which the Fefferman-Graham system of coordinates starts covering the AdS
space for a second time. This point corresponds to $\partial r /\partial \rho=0$, and 
results in the line $z_d^2(\rho)=4(1-H^2\rho^2)/(2H^2-H^4\rho^2-\mu)$ for
 $H^2>\mu$ and $z^2_d(\rho)=4(1-H^2\rho^2)/(\mu-H^4\rho^2)$ for $H^2<\mu$.
It is apparent from 
eq. (\ref{bb3a}) that the original variable $\phi$ cannot be
assumed to be periodic. This means that the metric (\ref{eqmetric4}) again represents (part of) the covering space of AdS.
The metric has a natural interepretation only for 
$\mu=0$, when it describes AdS space in Poincare coordinates (with $u=1/r$).

For $\mu\to 0^+$ we obtain
\begin{eqnarray}
r(z,\rho)&=&\frac{\sqrt{1-H^2\rho^2}}{z}+\frac{H^4\rho^2}{4\sqrt{1-H^2\rho^2}}z
\label{bbb2} \\
\phi(z,\rho)&=&\frac{1}{2H}\log\left[\frac{1+H\rho}{1-H\rho} \right]
-\frac{H^2\rho z^2}{2(1-H^2\rho^2+H^4\rho^2z^2/4)}.
\label{bbb3}
\end{eqnarray}
The minimal value $r_m(\rho)$ and the corresponding values of $z$ and $\phi$ are given by 
\begin{eqnarray}
z_m(\rho)&=&2\sqrt{\frac{1-H^2\rho^2}{H^4 \rho^2 }}
\label{zmrho} \\
r_m(\rho)&=&H^2 |\rho|
\label{rmrho} \\
\phi_m(\rho)&\equiv&\phi(z_m(\rho),\rho)=\frac{1}{2H}\log\left[\frac{1+H\rho}{1-H\rho} \right]
-\frac{1}{H^2\rho}.
\label{phimrho} 
\end{eqnarray}
The values at which the Fefferman-Graham coordinates start covering the AdS space for a second time are
\begin{eqnarray}
z_t(\rho)&=&2\sqrt{\frac{1-H^2\rho^2}{2H^2-H^4 \rho^2 }}
\label{zmrhod} \\
r_t(\rho)&=&r(z_t(\rho),\rho)=\frac{H}{\sqrt{2-H^2\rho^2}} 
\label{rmrhod} \\
\phi_t(\rho)&\equiv&\phi(z_t(\rho),\rho)=\frac{1}{2H}\log\left[\frac{1+H\rho}{1-H\rho} \right]
-\rho.
\label{phimrhod} 
\end{eqnarray}

As in the previous sections, the metric (\ref{eqmetric4}) joins parts of AdS space. The part of main interest 
starts from the AdS boundary at $z=0$. For given $t$ and $\rho$,  it is natural to assume that it
extends up to the point at which the Fefferman-Graham system starts covering the AdS space for a second time.
For variable $\rho$, the points generate the line of eq. (\ref{zmrhod}).
Notice that the covered space does not include the line (\ref{zmrho}), on which $r$ is maximized, as 
$z_t(\rho)<z_m(\rho)$. However, the two lines approach each other for $\rho \to \pm 1/H$, $z_m(\rho)\to 0$. The endpoints of
both lines are the horizons of de Sitter space. We can view the line segment $z_t(\rho)$ with  $-1/H< \rho < 0$ as the holographic 
dual of the horizon at $\rho=-1/H$, and the segment with  $z_t(\rho)$ with  $0< \rho < 1/H$ as the holographic 
dual of the horizon at $\rho=1/H$. It is important that, near $\rho=\pm 1/H$ where $z_t(\rho)\simeq z_m(\rho)$, 
these lines delimit the part of AdS space covered by the parametrization. A pictorial view of this behavior is given in 
ref. \cite{prepare}.

The AdS boundary is approached for a second time for $z\to \infty$. In constrast to the cases discussed earlier, the 
boundary metric does not have a simple form, as can be seen from eq. (\ref{eqmetric4}). 
This is also reflected in its holographic dual. The Fefferman-Graham
coordinates cover the boundary at $z\to \infty$ twice, while two AdS copies are employed in the near-boundary region.
A pictorial view of this construction is given in \cite{prepare}, while its interpretation is left for future work. In the 
following we concentrate on the CFT on the boundary at $z=0$.

We next turn to the calculation of the stress-energy tensor. 
For the metric of eq. (\ref{eqmetric4}) we obtain
\begin{eqnarray}
\label{eq3eeii} 
\rho&=&-\langle T_{~t}^{t} \rangle=
\frac{1}{16\pi G_3}\left(\frac{\mu-H^2}{1-H^2\rho^2}-H^2 \right)
\\
p& =& \langle T_{~\rho}^{\rho} \rangle = 
\frac{1}{16\pi G_3}\left(\frac{\mu-H^2}{1-H^2\rho^2}+H^2 \right),
\label{eq3epii}  
\end{eqnarray}
on a boundary with metric (\ref{eqmetric30}). Similarly to the Rindler case, 
we identify the ground state with the solution with $\mu=0$. 
The stress-energy tensor displays the expected singularities at the locations of the horizons $\rho=\pm1/H$ of static
de Sitter space \cite{bd}.
The conformal anomaly is
\begin{equation}
 \langle T_{~\mu}^{\mu\,(CFT)} \rangle  = 
\frac{1}{8\pi G_3}H^2.
\label{confaneii}\end{equation}

\subsection{Excited state}
\label{esgdb}
Similarly to the Rindler boundary, we can find a form of the metric that corresponds to an excited state of
the dual CFT. 
The metric (\ref{eqmetric}) with $\mu=0$ can be put in the form
\begin{equation}
\label{eqmetric3} 
ds^2
= \frac{1}{z^2} \left[ dz^2 +\left(1-\frac{1}{4}H^2 z^2 \right)^2 \left(
- (1-H^2\rho^2) d\eta^2 
+  \frac{d\rho^2}{1-H^2\rho^2}\right) \right],
\end{equation}
with a de Sitter boundary metric (\ref{eqmetric30}).
The coordinate transformation that achieves this is
\begin{eqnarray}
t(\eta,z,\rho)&=&\frac{1}{H}-\frac{1}{H}\frac{\exp[-H\eta]}{\sqrt{1-H^2\rho^2}}\frac{1+H^2z^2/4}{1-H^2z^2/4}
\label{aa1} \\
r(\eta,z,\rho)&=&\exp[H\eta]\sqrt{1-H^2\rho^2}\left(\frac{1}{z}-\frac{H^2}{4}z\right)
\label{aa2} \\
\phi(\eta,\rho)&=&\rho \frac{\exp[-H\eta]}{\sqrt{1-H^2\rho^2}}.
\label{aa3}
\end{eqnarray}

The stress-energy tensor can be computed from the above metric. It is 
\begin{eqnarray}
\label{eq3ee} 
\rho&=&-\langle T_{~t}^{t} \rangle=
-\frac{H^2}{16\pi G_3}\\
p& =& \langle T_{~\rho}^{\rho} \rangle = 
\frac{H^2}{16\pi G_3}.
\label{eq3ep}  
\end{eqnarray}
The conformal anomaly is given by eq. (\ref{confaneii}).
The above expressions can be reproduced by eqs. (\ref{eq3eeii}), (\ref{eq3epii}) in the presence of a positive
energy scale $\mu=H^2$. 

\section{Expanding boundary}
\label{expbound}
In this section we summarize the expressions for the case of 
a time-dependent boundary of the form
\begin{equation}
\label{eqmetricb} ds_0^2 = g_{\mu\nu}^{(0)} dx^\mu dx^\nu = 
-d\tau^2 + a^2(\tau) d\phi^2,
\end{equation}
with $a(\tau)$ an arbitrary function. A more detailed analysis is given in ref.
\cite{lamprou}.
The metric (\ref{eqmetric}) can be expressed as
\begin{equation}
\label{eqmetric35} 
ds^2= \frac{1}{z^2} \left[ dz^2 
- \mathcal{N}^2(\tau,z) d\tau^2 + \mathcal{A}^2 (\tau,z) d\phi^2 \right],
\end{equation}
with
\begin{eqnarray}
\label{aa}
\mathcal{A}(\tau,z) &=& a(\tau)\left(1+ \frac{\mu-\dot{a}^2(\tau)}{4a(\tau)^2}z^2\right) \\
\label{nn}
\mathcal{N}(\tau,z) &=&1- \frac{\mu-\dot{a}^2+2 a \ddot{a}}{4a^2}z^2
=\frac{\dot{\mathcal{A}}(\tau,z)}{\dot{a}}.
\end{eqnarray}
The above form is a generalization of the static metric (\ref{eqmetric1}) discussed in section \ref{mink}. The latter 
is reproduced from eq (\ref{eqmetric35}) for $a(\tau)=1$.

The comparison of eqs. (\ref{eqmetric1}), (\ref{eqmetric35})
indicates that 
\begin{equation}
 r(\tau,z) = \frac{\mathcal{A}(\tau,z)}{z}= \frac{a}{z} +\frac{\mu-\dot{a}^2}{4} \frac{z}{a}. 
\label{eqsys4} \end{equation}
The coordinate $\phi$ remains unaffected by the transformation. We assume that it is periodic, with periodicity $2\pi$.
As a result, the metric (\ref{eqmetric35}) is another representation of the BTZ black hole with mass $\sim \mu$. 
The dual picture is that of a thermalized CFT on an expanding background, with a scale factor $a(\tau)$. 
The mass $\sim\mu$ of the BTZ black hole sets the scale of the temperature.

The coordinates $(\tau,z)$ do not span the full BTZ geometry. As in the static case, they cover the two regions outside the 
event horizons,
located at
\begin{eqnarray}
\label{ze1}
 z_{e1}&=&\frac{2a}{\sqrt{\mu}+\dot{a}},
\\
\label{ze2}
z_{e2}&=&\frac{2a}{\sqrt{\mu}-\dot{a}}.
\end{eqnarray}
The quantities $z_{e1}$, $z_{e2}$ are the two roots of the equation
$r(\tau,z)=r_e=\sqrt{\mu}$. 
Moreover, the coordinates cover part of the regions behind the horizons. For constant $\tau$, the minimal value of $r(\tau,z)$ 
is obtained
for  
\begin{equation}
\label{zm}
z_m(\tau)=\frac{2a}{\sqrt{\mu-\dot{a}^2}},
\end{equation}
corresponding to 
\begin{equation}
r_m(\tau)=\sqrt{\mu-\dot{a}^2}.
\label{rm} \end{equation}
Clearly, $r_m\leq r_e$. For $\mu \leq \dot{a}^2$, the Fefferman-Graham parameters cover the whole
range of $r$, with $z$ taking values in the range $[0,2a/\sqrt{\dot{a}^2-\mu}]$.

The transformation of the time coordinate 
is more complicated. 
For $z <z_{e1}$, or for $z > z_{e2}$,
it is given by 
\begin{equation}
\label{tautz1}
t(\tau,z)=\frac{\epsilon}{2\sqrt{\mu}}\log\left[
\frac{4 a^2-\left(\sqrt{\mu}+\dot{a}\right)^2 z^2 }{ 4 a^2-\left(\sqrt{\mu}-\dot{a}\right)^2 z^2}
\right]+\epsilon\, c(\tau),
\end{equation}
where the function $c(\tau)$ satisfies $\dot{c}=1/a(\tau)$ and $\epsilon=\pm 1$.
For $z=0$ and $\epsilon = 1$ we have $t=c(\tau)$, where $c(\tau)$ is the conformal
time on the boundary.
For $z_{e1}<z<z_{e2}$, the transformation is
\begin{equation}
\label{tautz2}
t(\tau,z)=\frac{\epsilon}{2\sqrt{\mu}}\log\left[
\frac{-4 a^2+\left(\sqrt{\mu}+\dot{a}\right)^2 z^2 }{ 4 a^2-\left(\sqrt{\mu}-\dot{a}\right)^2 z^2}
\right]+\epsilon \, c(\tau).
\end{equation}
The values of $\epsilon$ may be chosen differently in the regions 
$z<z_{e1}$, $z_{e1}<z<z_{e2}$, $z>z_{e2}$.
 It is obvious from eq. (\ref{tautz1}), (\ref{tautz2}) that the transformation from
$(t,r)$ to $(\tau,z)$ coordinates is always singular on the event horizons. 

Similarly to the static case, the AdS boundary is approached in the limit $z\to 0$.
The thermalized CFT lives on a background that expands with a rate determined by $a(\tau)$. 
For $\mu > \dot{a}^2$, it is natural to assume that, at a given time $\tau$, the holographic dual is spanned by $z$ taking values 
in the interval $[0,z_m(\tau)]$.  Larger values of $z$ span the holographic dual of a CFT living on the AdS boundary approached 
for $z\to \infty$. Similarly to the previous section, we can define a coordinate $z'=1/z$ and obtain a Fefferman-Graham 
expansion around $z'=0$. The resulting boundary metric does not have a simple form, and we shall not discuss its
interpretation here.

The discussion of the construction with an expanding boundary in terms of global AdS coordinates is very complicated because of the 
identifications imposed by the BTZ construction and the singular transformation (\ref{tautz1}), (\ref{tautz2}). Moreover, 
the point of view of a boundary observer is best expressed in terms of the
time coordinate $\tau$. For this reason we base the discussion of the entropy for an expanding boundary on the 
$(\tau, z)$ coordinates.

The stress-energy tensor of the dual CFT on the time-dependent boundary 
is determined via holographic renormalization and is given by eq.~(\ref{eq3a}). 
We obtain the energy density and pressure, respectively,
\begin{eqnarray}
\label{eq3te} 
\rho&=&\frac{E}{V}=-\langle T_{~\tau}^{\tau} \rangle=
\frac{1}{16\pi G_3}\frac{\mu-\dot{a}^2}{a^2}
\\
P& =& \langle T_{~\phi}^{\phi} \rangle = 
\frac{1}{16\pi G_3}\frac{\mu-\dot{a}^2+2a\ddot{a}}{a^2},
\label{eq3tp}  
\end{eqnarray}
on a boundary with metric 
(\ref{eqmetricb}).
We deduce the conformal anomaly
\begin{equation}
 \langle T_{~\mu}^{\mu\,(CFT)} \rangle  = 
\frac{1}{8\pi G_3}\frac{\ddot{a}}{a}.
\label{confan}\end{equation}

As a final comment in this section, we note that a wormhole construction with a static throat that does not coincide with 
the horizon has been proposed in ref. \cite{solodukhin} as a framework for restoring unitarity in the
relaxation of the BTZ black hole. Contrary to the metrics we described above and in section \ref{mink}, the proposed 
configuration is not a solution of the Einstein equations.

\section{Length of the throat and entropy}
\label{lthr}
All the coordinate systems for the (2+1)-dimensional AdS space and the BTZ black hole that we discussed
in the previous sections have the Fefferman-Graham form (\ref{eq2}), (\ref{fefg}).
The metric on the boundary at $z=0$, for which we have chosen several nontrivial forms,
 is given by the first term in this expansion. The system of coordinates (\ref{eq2}) is adapted to a
boundary observer. It provides a natural interpretation of the bulk geometry in the dual picture,  as 
exemplified by the form of 
the stress-energy tensor of the dual CFT (\ref{eq3a}).  In this section we suggest that 
the entropy associated with the dual theory on a nontrivial background is related to the form of the
bulk metric in this coordinate system.

The common feature 
of the various forms of the bulk metric that we presented in the previous sections is that the 
coordinates do not span the full bulk geometry. At any given time, the metric describes two asymptotic regions 
(in the limits $z\to 0$ and $z\to \infty$), joined by a two-dimensional surface. If the spatial coordinate of the 
boundary geometry is periodic, as when the bulk geometry is that of the BTZ black hole, the surface has the form of a throat. 
In the opposite case, the bulk is the covering space of AdS and the two-dimensional surface  is better characterized
as a bridge. The narrowest part of the throat or bridge is a line that defines the boundary of the
part of the bulk geometry that is not covered by the Fefferman-Graham parametrization. In a sense, it 
determines the part of the bulk that is not included in the construction of the dual theory. 
In the context of the AdS/CFT correspondence it seems natural to attribute the entropy of the dual theory to
the presence of this line. In more quantitative terms, the total entropy may be assumed to be proportional to
the length $A$ of the line. In the following, we shall test whether such a conjecture gives reasonable predictions for the 
various boundary metrics that we discussed in the previous sections. 

As we discussed at the end of section \ref{mink}, the boundary of the AdS space is approached in 
both limits $z\to 0$ and $z\to \infty$. As a result, the Fefferman-Graham metric in 2+1 dimensions 
encodes information on the dual CFT on two different backgrounds. Our construction generates
standard boundary metrics in the $z\to 0$ limit, while the boundary metrics for $z \to \infty$ may be unconventional.
For this reason our analysis focuses on the CFT living on the $z=0$ boundary. We assume that the holographic
dual is provided by the part of the AdS space between the $z=0$ boundary and the line corresponding to the narrowest 
part of the throat.

Before proceeding, we need to determine the proportionality factor $k$ between the entropy and the 
length $A$. We assume $k$ to be a universal constant. 
In the case of a flat boundary, discussed in section \ref{mink},
a nonzero value of $\mu$ for the bulk metric corresponds to a thermalized CFT at
a temperature equal to that of the BTZ black hole: $T_{th}=\sqrt{\mu}/(2\pi)$. 
The narrowest part of the throat described by the metric of eq. (\ref{eqmetric1}) coincides with
the event horizon of the black hole. It is located at $z=2/\sqrt{\mu}$, or $r=\sqrt{\mu}$.
Its length is given by $\sqrt{\mu} V$, where $V$ is the length of the space spanned by $\phi$, equal to $2\pi$ for the BTZ
black hole. For a Minkowski boundary, the CFT entropy coincides with the entropy of the black hole, given by the last of 
eqs. (\ref{eqTM}). 
This allows us to determine that $k=1/(4G_3)$. Our conjecture about the entropy for a general boundary metric becomes
\be
S= \frac{1}{4G_3} A,
\label{entropyy} \ee
with $A$ the length of the narrowest part of the throat or bridge at a given time.

\subsection{Rindler entropy}
\label{rindentr}

The Euclidean version of the two-dimensional Rindler metric (\ref{rind})
has a conical singularity at $x=0$, unless the Euclidean time is assumed to be periodic with 
period $2\pi/a$.  This indicates that the space has an intrinsic temperature, given by 
$T_R=a/(2\pi)$. The entropy of the dual theory is expected to be nonzero even in the
absence of a black hole in the bulk geometry. 

For the calculation of the ground-state entropy we consider the metric (\ref{rindb}) with $\mu= 0$. 
The situation is static, so that the calculation of the entropy can be performed at any value of the time coordinate $t$. 
We select $t=0$ for convenience.
As we saw in subsection \ref{gsrindd}, 
the minimal value of $r$, for fixed $x$, is obtained for $z_m(x)$ given by eq. (\ref{zmx}). 
This value is constant, given by eq. (\ref{rmx}), and  defines a line corresponding to the narrowest part of the 
bridge.  Its length can be obtained through the integration of the line element (\ref{rindb}) for $dt=0$. 
However, it is easier to perform the calculation in the original $(r,\phi)$  system of coordinates.
As the change of coordinates of eqs. (\ref{brind2}), (\ref{brind3a}) does not involve the
time coordinate, we can employ the line element 
(\ref{eqmetric}) with $dt=0$. Using $r=r_m(x)=a$ and $\phi=\phi_m(x)$ given by eq. (\ref{phimx}), we find that the
entropy (\ref{entropyy}) is given by 
\be
S=\frac{1}{4G_3}  \int_{-\infty}^{\infty} a d\phi = \frac{1}{4 G_3} \int_0^\infty \frac{dx}{x} = 
 \frac{1}{4 G_3} \int_0^\infty \frac{dz}{z} .
\label{srind} \ee
All versions of the integral are plagued by infinities.
The expression for the entropy diverges and its meaning is not clear.

The way to handle this situation is to introduce a cutoff $\epsilon$ for 
the line element, with $\epsilon \ll 1$. The dominant contribution to the integral of eq. (\ref{srind}) can be written
as 
\be
S=\frac{2}{4 G_3} \int_{\epsilon} \frac{dz}{z},
\label{regulr} \ee
with the factor of 2 arising from the two limits $\phi\to\pm \infty$.
We have not introduced an explicit upper limit, as its exact value is
irrelevant for $\epsilon \to 0$. 

In order to extract the physical meaning of eq. (\ref{regulr}), we must compare it with the effective Newton's constant $G_2$
of the (1+1)-dimensional theory. In the context of the AdS/CFT correspondence the boundary gravity is not dynamical and
$G_2$ vanishes. The boundary gravity becomes dynamical in the context of the Randall-Sundrum model \cite{rs}, which
employs only a part of the full AdS space. For the (2+1)-dimensional case that we are considering we have 
\be
\frac{1}{G_2}=\frac{1}{G_3}\int_{z_1}^{z_2}\frac{dz}{z}, 
\label{rsgn} \ee
where $z_1$, $z_2$ denote the locations of the two branes. In the previous
sections we saw that the construction with a nontrivial conformal boundary leads to the presence of two
dual theories, living at $z\to 0$ and $z\to \infty$. Our interest lies in the $z=0$ theory. A consistent holographic
description can be obtained if we restrict the range of $z$ within distances from the boundary such that the Fefferman-Graham
coordinates cover the (part of) AdS space only once. 
For the effective Newton's constant we should then use 
the lower limit $z_1$ as a regulator: $z_1=\epsilon$, with $z_2$ kept at a finite value.  This gives
\be
\frac{1}{G_2}=\frac{1}{G_3}\int_{\epsilon}\frac{dz}{z}, 
\label{rsgn2} \ee
where we have omitted the upper limit, as the integral is dominated by the lower one for $\epsilon \to 0$.
The fact that $G_2$ tends to zero in the same limit is consistent with the absence of dynamics for the 
boundary gravity in the context of AdS/CFT.

Comparison of eqs. (\ref{regulr}), (\ref{rsgn2}) indicates that we can interpret eq. (\ref{srind}) as
\be
S=\frac{2}{4G_2}.
\label{srindd} \ee
Repeating the calculation of ref. \cite{laflamme} for the two-dimensional case results in an expression 
for the entropy of the two-dimensional Rindler wedge that is half the value given by eq. (\ref{srindd}). 
The reason for the discrepancy becomes apparent if we consider the 
relations between the coordinates $x$, $\phi$, $\phit$ on the line with $r=a$.
These are given by  
eq. (\ref{phimx}) and  eq. (\ref{globalf}) with $r=a$ and $t=0$. 
The compactness of the spatial direction on the AdS boundary implies that 
the global coordinate $\phit$ is periodic.
As a result, in the construction with a Rindler conformal boundary, the limits $x\to 0$ and $x \to \infty$ of the 
Fefferman-Graham coordinate $x$ must be identified. In terms of global coordinates, they both correspond to the
point $\phit=\chit=\pi/2$ on the AdS boundary. 
A pictorial representation of this feature is given in ref. \cite{prepare}.

We are thus led to the conclusion that the construction of section \ref{rindd} does not represent the conventional Rindler
space. It provides a holographic description of the Rindler wedge, with an identification of the limits $x\to 0$ and $x\to \infty$. 
Completing the space trivially by considering a second copy of AdS space for the wedge with $x<0$ does not 
resolve this problem, as the identification of the limits $x\to 0^{\pm}$ and $x\to \pm \infty$ will remain. 

Despite these misgivings, the region near $x=0$ provides an acceptable holographic description of the 
Rindler horizon. The defining feature is the absence of complete coverage of the $t=0$ slice of AdS space. The
entropy can be identified with the length of the boundary of the excluded region near $x=0$. This is dominated by the
part of the line near the AdS boundary, and is given by 
\be
S_R= \frac{1}{4G_2}.
\label{srinddd} \ee
The extension of this calculation for different Euclidean-time slices is given in ref. \cite{prepare}.
The conclusion remains the same. At any time, the line delimiting the region covered by the Fefferman-Graham
parametrization in the vicinity of $x=0$ has a length that reproduces eq. (\ref{srinddd}).

\subsection{de Sitter entropy}
\label{desientr}

The case of a de Sitter boundary is very similar to the Rindler case. The space has a characteristic temperature
$T_{dS}=1/H$. The part of AdS space covered by the Fefferman-Graham coordinates is delimited by the line 
defined through eqs. (\ref{zmrhod})-(\ref{phimrhod}). This line approaches the AdS boundary for 
$\rho\to \pm 1/H$, where its length diverges. When expressed in terms of global coordinates, both these limits can be 
seen to correspond to the point $\phit=\chit=\pi/2$ on the AdS boundary. 
The identification of the two limits, imposed by the periodicity of the global
coordinate $\phit$, does not alter the physical picture, as they correspond to the 
two symmetric horizons of the (1+1)-dimensional de Sitter space. 
The horizons are the holographic images 
of the line segments delimiting the region covered  by the Fefferman-Graham 
coordinates on either side of the point $\phit=\chit=\pi/2$.

The sum of the (infinite) lengths of the two segments ending at the point  $\phit=\chit=\pi/2$
is related to the de Sitter entropy. 
After regularization, the entropy is given by eq. (\ref{regulr}), which leads to 
\be
S_{dS}=\frac{1}{2G_2}.
\label{srdes} \ee
This is the correct expression for the entropy of two-dimensional de Sitter space \cite{dsen}.

\subsection{CFT Entropy on an expanding background }
\label{frwentro}
We now turn to the case of an expanding boundary with a scale factor $a(\tau)$ that is an arbitrary function of time.
In this case the bulk geometry is that of a BTZ black hole, whose mass parameter sets the scale for the 
thermal energy of the CFT in the dual picture. 
We saw in section \ref{expbound} that the Fefferman-Graham coordinates cover a part of the geometry that reaches
behind the event horizon of the BTZ black hole. The minimal value of $r$ is $r_m(\tau)=\sqrt{\mu-\dot{a}^2}$, which is
smaller than the value for the event horizon $r_e=\sqrt{\mu}$. 

The identification of the entropy can be made in two different ways. One could adopt the point of view that the
event horizon remains the relevant surface (line in the BTZ case) for the definition of the entropy, even for a 
time-dependent boundary. This would mean that the CFT entropy remains constant  during the expansion and the
entropy density scales $\sim 1/a$. The consistency of this assumption would require that the temperature scale
$\sim 1/a$ and the thermal energy density $\sim 1/a^2$. The energy density of eq. (\ref{eq3te}) contains two terms. 
The one $\sim \mu/a^2$ should be identified with the thermal energy density of the dual CFT. The term $\sim \dot{a}^2/a^2$
would then be interpreted as time-dependent vacuum energy, which does not have any counterpart in the entropy.

The alternative point of view would be to identify the entropy with the length of the line corresponding to the 
narrowest part of the throat at a given time $\tau$. The location of this line is given by eq. (\ref{rm}) and its length
is $2\pi\sqrt{\mu-\dot{a}^2}$.  Our prescription (\ref{entropyy}) would then give 
\be
\label{entropyfrw}
S=\frac{\pi}{2 G_3}\sqrt{\mu-\dot{a}^2}.
\ee

In order to obtain some intuition on this point, it is instructive to rederive the entropy associated with the
BTZ black hole using the Cardy formula \cite{cardy}.
It has been shown \cite{henneaux} that the asymptotic symmetries of (2+1)-dimensional Einstein 
gravity with a negative cosmological constant correspond to a pair of Virasoro algebras, with 
central charges $c=\ct=3/(2 G_3)$. For the case of the BTZ black hole, 
the eigenvalues $\Delta$, $\Dt$ of the generators $L_0$, $\Lt_0$ of these 
algebras can be calculated in terms of the BH mass \cite{banados,carlip2}. They are
\be
\Delta=\Dt=\frac{\mu}{16 G_3}.
\label{eigenv} \ee
The Cardy formula allows the calculation of the density of states and, therefore, the associate entropy.
It gives
\be
S=2\pi\sqrt{\frac{c}{6}\left( \Delta-\frac{c}{24}\right)} +2\pi\sqrt{\frac{\ct}{6}\left( \Dt-\frac{\ct}{24}\right) } =
\frac{\pi}{2G_3}\sqrt{\mu-1}.
\label{cardyent} \ee
In the limit $\mu \gg 1$, this expression reproduces correctly the BH entropy (third of eqs. (\ref{eqTM}) with $V=2 \pi$), 
which we identify with the CFT entropy. 

The crucial element in eq. (\ref{cardyent}) is the shift of the mass term by 1. This reflects the 
Casimir energy resulting from the compactness of the angular coordinate $\phi$. The Casimir energy 
is apparent in the holographic stress-energy tensor (\ref{eq3e}) of the theory dual to the metric (\ref{eqmetric}) with
$\mu=-1$. As we have mentioned in section \ref{3db}, for $\mu=-1$ eq. (\ref{eqmetric}) describes AdS space 
in global coordinates, which can be viewed as the ground state (no thermal energy). It is clear then that the Casimir energy affects the
entropy. This observation has led to the proposal of ref. \cite{verlinde} (the so-called Cardy-Verlinde formula), in which
all the central charges appearing in eq. (\ref{cardyent}) are replaced by the Casimir energy. The conjecture is put forward
that the resulting formula gives the entropy for time-dependent backgrounds in dimensions higher than two.

Our result (\ref{entropyfrw}) gives a different generalization of the
Cardy formula in two dimensions. The total energy density on a time-dependent bacground is given by 
eq. (\ref{eq3te}). The contribution $\sim \dot{a}^2$ results from the vacuum energy generated by 
the curvature related to the expansion. For $\mu\gg 1$, the effect is to replace $\mu$ by $\mu-\dot{a}^2$ 
in the expression for the energy.
Making this replacement in eq. (\ref{cardyent}) exactly reproduces our result (\ref{entropyfrw}).

The result (\ref{entropyfrw}) has some intriguing implications. For a decelerating expansion, 
with $\dot{a}(\tau)$ diminishing with time, the total entropy increases and asymptotically approaches the
value for the static case. For an accelerating expansion the entropy decreases until it vanishes for $\dot{a}^2=\mu$. 
This is a sign of instability of the system, or that the corresponding state is unphysical. A very similar situation
was encountered in the study of the CFT entropy on a four-dimensional FRW background 
in terms of the (4+1)-dimensional AdS black hole \cite{tetradis}.
In that case the instability is associated with the deconfining phase of the 
$\mathcal{N}=4$ super Yang-Mills theory on a background with accelerating expansion. 
The interpretation of the two-dimensional case we are considering depends on the embedding in
a dynamical theory (with additional fields). 

\section{Conclusions}
\label{conclu}

The purpose of this work was to identify information that can be extracted through the AdS/CFT correspondence
from a bulk geometry that is asymptotically AdS but has a nontrivial boundary. The (2+1)-dimensional AdS space
provides the simplest setting for such a study. Expressed in appropriate coordinates,
it can describe ground and excited states of the dual CFT.
Moreover, through appropriate identifications it can be turned into the BTZ black hole \cite{btz1,btz2}, thus allowing the 
study of the properties of CFTs at nonzero temperature. 
We constructed metrics that 
have as boundaries Minkowski, Rindler, static de Sitter and FRW two-dimensional spaces. 
All this was achieved through the use of appropriate Fefferman-Graham coordinates \cite{fg,Skenderis}. 
Most of the forms of the metric describe the ground state of the dual CFT on the respective background, but
we have also constructed the dual metrics of excited states. All these metrics are locally isometric to AdS, but
have completely different physical interpretations in the dual picture. 
We have computed the stress-energy tensor for all these solutions through holographic renormalization \cite{Skenderis}.
It has the expected form for each two-dimensional background. 

The focus of the study was on the CFT entropy on the various backgrounds. In this context the Feffermann-Graham
system of coordinates plays a special role, as it is adapted to a boundary observer. 
We have found that, in general, 
this system does not cover the whole bulk geometry. A throat or bridge appears
that defines the boundary of the region covered by the Fefferman-Graham parametrization.
We proposed that the narrowest part of the throat or bridge can be identified with the entropy of the dual CFT on 
the corresponding background. The exact expression is given by eq. (\ref{entropyy}), with $A$ the length of the 
narrowest part at a given time. 

We first tested this proposal for Rindler and de Sitter boundaries.
Through an appropriate regularization of the near-boundary bulk space-time, we
showed that the entropy of two-dimensional Rindler and de Sitter spaces is reproduced correctly. 
Then we
considered a boundary of the FRW type, which generates a time-dependent throat. 
In this case our proposal results in a generalization of the Cardy formula \cite{cardy} for the entropy of
the two-dimensional CFT. The resulting entropy takes into account the Casimir energy generated by the
curvature related to the expansion of the bacground.

Our construction bears strong resemblance to holographic calculations of the entanglement entropy of 
the dual CFT within a specified region ${\cal A}$ on the boundary \cite{ryu,takayanagi,hubeny}. 
In these works the entropy is identified with the area of 
an appropriately defined minimal surface ${\cal S}$ that starts from the border of ${\cal A}$ on the boundary and 
extends into the bulk. When ${\cal A}$ is taken to cover the whole boundary,  ${\cal S}$ tends to wrap around
obstructions in the bulk space-time, such as black-hole horizons. In this case it reproduces the thermodynamic entropy of the 
dual CFT. 
In the studies of refs.  \cite{ryu,takayanagi,hubeny} the boundary is assumed to be flat, even when time dependence is 
introduced in the bulk metric (such as in the case of a bulk Vaidya-AdS space).
We expect that, for nontrivial backgrounds such as the ones we considered, the two proposals would agree, as
the minimal surface  ${\cal A}$ would wrap around the throats or bridges of our construction.

Our proposal can be generalized to higher-dimensional spaces. The technical difficulty of constructing appropriate forms of the 
bulk metric is expected to increase with the number of dimensions. Work in this direction is in progress.

\section*{Acknowledgments}
I wish to thank T. Christodoulakis, G. Diamandis, E. Kiritsis, I. Papadimitriou, K. Skenderis, P. Terzis
for useful discussions. 
This work was supported in part by the EU Marie Curie Network ``UniverseNet'' 
(MRTN--CT--2006--035863) and the ITN network
``UNILHC'' (PITN-GA-2009-237920).

\end{document}